\documentclass[12pt,twocolumn]{aastex631}
\begin{document}

\title{ROME IV. An Arecibo Search for Substellar Magnetospheric Radio Emissions in Purported Exoplanet-Hosting Systems at 5 GHz}

\author[0000-0001-6987-6527]{Matthew Route}
\affiliation{Department of Astronomy and Astrophysics, Pennsylvania State University, 525 Davey Laboratory, University Park, PA 16802, USA}
\affiliation{Center for Exoplanets and Habitable Worlds, Pennsylvania State University, 525 Davey Laboratory, University Park, PA 16802, USA}
\affiliation{Northrop Grumman Electronic Systems, 6120 Longbow Drive, Boulder, CO 80301, USA}
\affiliation{Department of Physics and Astronomy, University of Mississippi, 121B Lewis Hall, University, MS 38677, USA}

\correspondingauthor{Matthew Route}
\email{mproute@olemiss.edu}


\begin{abstract}
Plasma flow-obstacle interactions, such as those between an exoplanet's magnetosphere and the host star's stellar wind, may lead to detectable radio emissions.  Despite many attempts to detect magnetospheric (auroral) radio emissions from exoplanets, a reproducible, unambiguous detection remains elusive.  This fourth paper of the ROME (Radio Observations of Magnetized Exoplanets) series presents the results of a targeted radio survey of nine nearby systems that host exoplanet, brown dwarf, or low-mass-stellar companions conducted with the Arecibo radio telescope at $\sim$5 GHz.  This search for magnetospheric radio emissions has the greatest sensitivity ($\sim$1 mJy during $<$1 s integration times) and collected full Stokes parameters over the largest simultaneous bandpass of any survey to date.  It is also the first survey to search for radio emission from brown dwarfs of spectral class Y, which may illuminate open questions regarding their magnetism, interior and atmospheric structure, and formation histories.  No magnetospheric radio emissions from substellar companions were detected.  These results are examined within the context of recent theoretical work on plasma flow-obstacle interactions, and radio emissions observed from the solar system planets and ultracool dwarfs.
\end{abstract}
 
\section{Introduction}

Despite a long history of searching for magnetospheric radio emissions from exoplanets, a lack of reproducible detections leaves the potential scientific value of the discipline yet to be realized.  Nevertheless, its importance to understanding the magnetic properties and related processes of exoplanets guarantees that it should continue.  The electrodynamic engine for these emissions, commonly known as (exo)planetary auroral emissions, is primarily thought to involve the dissipation of kinetic flow power and/or Poynting flux from a weakly magnetized stellar wind (plasma flow) across a (exo)planetary magnetospheric cross-section (an obstacle with a strong magnetic field).  Some of the dissipated solar wind energy accelerates electrons to $\gtrsim$keV energies, a fraction of which, in turn, is converted into auroral emissions that span the electromagnetic spectrum from radio to X-ray wavelengths.  The observational signatures of auroral radio emissions have been well characterized for the magnetized solar system planets \citep{zar98,bad15,zar18}.  An additional potential electrodynamic engine is magnetosphere-ionosphere coupling currents that arise from the departure of rigid corotation of exomoon plasma within (exo)planetary magnetospheres (e.g., \citealt{nic20}).  This work focuses on stellar wind-(exo)planetary magnetospheric interactions due to their prevalence within the solar system, and their dependence on observed exoplanet properties that may be used as target selection criteria.

The importance of exoplanetary radio emissions to modern astrophysics rests with their ability to serve as a tool to discover new exoplanets (e.g., \citealt{laz18}), diagnose (exo)planet interior structure, composition, and thermodynamic properties (e.g., \citealt{chr09,yan21}), properties of the ionosphere and surrounding plasma environment (e.g., \citealt{nic20}), atmospheric erosion (e.g., \citealt{gri15,wor16}), membership and properties of orbiting (exo)moons (e.g., \citealt{zarplus18,car21}), and assess the habitability of the (exo)planets themselves or their orbiting (exo)moons (e.g., \citealt{gre21}).  Although several putative detections have been announced (e.g., \citealt{tur21} for $\tau$ Bo\"{o}tis and references therein), an unambiguous, reproducible discovery remains elusive.

Several decades of remote and \emph{in situ} observations of the Jovian magnetosphere have revealed that Jupiter provides an intriguing case study for magnetosphere‐ionosphere coupling, magnetic interactions with the Galilean satellites Io, Europa, and Ganymede, and to a lesser extent, magnetospheric interactions with the solar wind.  These interactions give rise to a legion of radio emissions observed from $<$1 kHz to $\gtrsim$20 GHz that include emissions $\lesssim$40 MHz generated by the electron cyclotron maser (ECM) operating in its polar regions, and higher-frequency emissions from synchrotron radiation located at its magnetic equator and poles.  The most powerful Jovian radio emissions, including the decametric (DAM) and rapid S-bursts, are caused by coherent ECM radiation that is $\sim$10$^{5}$ more luminous than the weakest emission, incoherent decimetric (DIM) synchrotron emission \citep{zar04,tre06,dep15,zarplus18,nic20,lou21}.  It is therefore prudent that any search for exoplanetary magnetospheric radio emissions should focus on detecting those generated by the ECM mechanism.

Beyond the solar system, \citet{yan77} sought to detect the first extrasolar planets via their magnetospheric radio emissions using Jupiter as a template (e.g., \citealt{zar98}).  This survey, and its numerous successors, failed to identify reproducible candidates (for a review, see \citealt{laz18}).  Many works sought to estimate the flux densities and frequencies of emissions from exoplanet magnetospheric interactions (e.g., \citealt{far99,laz04,gri07,ign10,rei10,nic12,gri17,zag18}).  Important ingredients in estimating the radio flux density are properties of the stellar wind (density, velocity, and interplanetary magnetic field), while exoplanet age, mass, and rotation rate determine the exoplanet magnetic field strength (e.g., \citealt{chr10}). Since, these scaling laws generally led to estimates of $B\sim$10-100 G magnetic fields, surveys sought exoplanet radio emission at low frequencies.  However, many of the theoretical scaling laws are based on the extrapolation of the radio emissions from $\leq$1 M$_{J}$ magnetized solar system planets to more massive exoplanets.  It is unclear whether this extrapolation is appropriate.  Indeed, the discovery of ``inflated'' hot Jupiters may imply that these exoplanets have stronger than predicted global magnetic fields, perhaps with dipole field strengths $B\lesssim$250 G.  Small-scale exoplanet fields may be even stronger (e.g., \citealt{yad17}).  Indeed, \citet{con22} determined small-scale, polar Jovian magnetic fields $B\sim$20 G, far in excess of the magnetic field strength measured by the Jovian dipole moment alone, $B$=4.18 G.  Given the historic difficulty of conducting sensitive low-frequency surveys, the evolution in exoplanet dynamo theories, the failure to detect exoplanet magnetospheric interactions, and the surprising discovery of powerful, $\sim$kG magnetic fields among ultracool dwarfs (UCDs), surveys at $\sim$GHz frequencies have been conducted as well (e.g., \citealt{win86,bas00,laz10,str12,rou13,sir14,har16,per21,cen22}).

While the magnetic properties of the solar system planets may anchor expectations of giant exoplanet magnetism at the lower mass end, UCDs, or low-mass stars and brown dwarfs of spectral types $\geq$M7, are of special interest because their radio emissions anchor the high-mass end, and overlap stellar and planetary regimes.  If not from intrinsic stellar-like flaring, their emissions stem from some type of plasma flow-obstacle interactions, although it is unclear if their electrodynamic engines consist of magnetosphere-ionosphere coupling, magnetic (unipolar or dipolar) interactions with orbiting bodies, or magnetospheric interactions with the interstellar medium (e.g., \citealt{sch09,wil14,hal15}).  Low-mass stars of spectral types $\gtrsim$M4, brown dwarfs, and massive exoplanets have fully convective interiors that generate their magnetic dynamos  \citep{chab00,wah17}.  Since cooler brown dwarfs have properties intermediate between stars and planets, it was widely assumed that their magnetic activity and field strengths would also be intermediate in nature.

Magnetic field evolutionary tracks that extrapolate from a scaling law derived from geodynamo models estimate that M dwarf, brown dwarf, and exoplanet magnetic field strengths peak at $\sim$4 kG, $\sim$3 kG, and $\sim$200 G, respectively, for objects of ages $\sim$1 Gyr \citep{chr09,rei10}.  During my first $\sim$5 GHz survey of UCDs conducted at Arecibo Observatory (AO), I detected three high brightness temperature ($T_{B}>10^{11}$ K), highly circular polarized ($\gtrsim$70\%) flares from the first radio-emitting T dwarf, 2MASS J10475385+2124234 (J1047+21; \citealt{rou12,rou13}).  The flare properties indicated that they were caused by the same ECM mechanism that operates within the Jovian and solar magnetospheres.  Application of the cyclotron formula to these flares revealed that this $\sim$900 K, T6.5 brown dwarf hosts a magnetic field $B\sim$1.7 kG.  Additional observations of J1047+21 at 10 to 15.75 GHz suggested that it may host magnetic fields $B\sim$5.6 kG, assuming ECM radiation at the fundamental frequency \citep{wb15,kao18}.  The later detection of five 15-100\% circularly polarized bursts of $T_{B}>4\times10^{11}$ K emission from the T6 brown dwarf WISEPC J112254.73+255021.5 (J1122+25) during my second 5GHz AO radio survey indicated that strong magnetic fields may be quite common among cool brown dwarfs with exoplanet-like effective temperatures \citep{rou16a,rou16b}.  Furthermore, the T2.5 brown dwarf SIMP 01365662+0933473, was estimated to be an $\sim$200 Myr substellar object of only $\sim$13 M$_{J}$, yet with a magnetic field strength $B\sim$3.2 kG \citep{kao18}.  Thus, the magnetic energy that powers the radio emissions from brown dwarfs with exoplanet-like effective temperatures exceeds the geodynamo scaling law estimates by approximately an order of magnitude \citep{kao18}.
 
Perhaps T dwarf magnetism as manifested at radio wavelengths may provide better insight as to how giant exoplanet magnetospheres behave than scaling laws extrapolated from $\leq$1 M$_{J}$ mass magnetized planets in the solar system.  The surprisingly strong magnetic field of J1047+21 in apparent violation of dynamo scaling laws suggests that giant exoplanets may also host magnetic fields stronger than anticipated that could generate radio activity at $\sim$GHz frequencies. Therefore, I surveyed nearby star systems with known substellar companions at AO in 2010-2011 to search for magnetospheric emissions from their substellar companions.  This work complements my survey of stellar radio emissions resulting from star-planet interactions (ROME III, \citealt{rou23}), and my investigation of multiwavelength observations of putative star-planet interactions within the HD 189733 system (ROME I and II; \citealt{rou19,rl19}).  Section 2 describes my target selection criteria, instrumentation, and observations.  Section 3 presents the resulting detection limits from this survey.  Section 4 contextualizes these results with respect to other surveys of exoplanet radio emissions, UCD radio activity, and recent theoretical work.  This section also describes the implications of these results on the search for exomoons orbiting exoplanets within their respective habitable zones (HZs).  Section 5 concludes by reviewing the significance of this survey and providing suggestions for future work.  The results of this survey indicate that the detection and characterization of exoplanet magnetism via magnetospheric emissions remains an admirable, although difficult, goal.

\section{Target Selection and Observations}
\subsection{Target Selection}

Target systems were selected for this radio survey based on the anticipated ability of their substellar (brown dwarf and exoplanet) companions to generate magnetospheric radio emissions in accordance with plasma flow- obstacle interaction theory.  In each case, targets were chosen based on the physical properties of the companions (i.e., mass, semimajor axis) and their location in the sky (Table 1).  Theoretical models suggest that the global magnetic field strength of an object is determined by the internal energy available to power its convective magnetic dynamo.  This internal energy can be estimated from parameters such as mass, radius, age, rotation rate, and internal composition and conductivity.  Substellar objects with larger masses have greater thermal fluxes to power convection-driven magnetic dynamos, resulting in larger mean magnetic field strengths \citep{chr09, rei10}.  Within the plasma flow- obstacle paradigm, more massive substellar objects have stronger magnetic fields that form larger magnetospheres.  These magnetospheres present larger obstacles to dissipate stellar wind kinetic and/or magnetic power, which should lead to greater radio luminosities \citep{zar18}.

The maximum frequency of ECM radio emissions is related to the local cyclotron frequency, $\nu_{c}$, by $\nu_{c}$[MHz] $= 2.8 n B$ [Gauss], where $n$ is the harmonic number, and $B$ is the local magnetic field strength determined by the internal energy available to power the convective magnetic dynamo.  It is apparent from this formula that more massive substellar objects with stronger magnetic fields will cause higher frequency emissions, perhaps even extending into the $\sim$GHz frequency range.

When a substellar object orbiting its host star is detected, it is customary to report its minimum mass, $M~sin~i$. However, in many cases, the uncertainty in its orbital inclination with respect to our line of sight makes it ambiguous whether the object is truly an exoplanet or is a brown dwarf.  Statistical analysis indicates that the true mass of a substellar companion is typically $\sim$15\% higher\footnote{ ``Exoplanets Data Explorer $|$ Table,'' available at http://exoplanets.org/table} (e.g., \citealt{ref11}).  Therefore, HD 38529 b (now HD 38529 c) and HD 114762 Ab were selected as candidates that straddle the exoplanet-brown dwarf boundary, since their reported masses were 12.7 and 11 M$_{J}$, respectively \citep{lat89,fis03}.  The radio emissions from these objects could be as strong as the $\sim$4-5 GHz emission from the T6.5 brown dwarf J1047+21.  Thus, putative planetary systems with relatively massive substellar companions were selected.

A second target selection criterion was that massive targets with smaller semimajor axes were preferred because they are hypothesized to have greater radio luminosities. Returning to the plasma flow- obstacle paradigm, substellar objects closer to their host stars have smaller magnetospheres due to the pressure balance between the object's magnetic field and the stellar wind ram pressure.  Although these magnetospheres would present smaller obstacles to the stellar wind plasma flow, as a stellar wind travels outward from the host star, mass and magnetic flux conservation dictate that objects with smaller semimajor axes will encounter greater kinetic flow powers and Poynting fluxes.  Therefore, close-in, massive substellar objects may present the most promising targets for magnetospheric emissions \citep{zar18}.

Intrinsic system properties aside, I targeted systems $<$50 pc from the Sun because even the most optimistic magnetospheric radio flux estimates (e.g., \citealt{laz04}) suggest only marginal detectability for targets observed at an appropriate frequency.  Additionally, the single, fixed-dish nature of the Arecibo radio telescope constrained the targets to have declinations of 0$^\circ$ to +38$^\circ$.

\subsection{Instrumentation and Observations}

The substellar companion hosting systems were surveyed from 2010 January 6 to 2011 July 20 under AO observing program A2471.  Each system was observed for a total of 0.67 to 2.50 hours, sometimes across multiple epochs, although each individual epoch was $<$2 hrs (Table 2\footnote{Data sets referenced in this table may be requested by submitting a ticket with the category ``Arecibo Data'' to the Arecibo Observatory Tape Library hosted by Texas Advanced Computing Center (https://www.tacc.utexas.edu/about/help/).}).  This duration was set by the maximum time targets can take to transit the fixed dish of the Arecibo radio telescope.  During the observing sessions, each 10-minute on-target science scan was bracketed by 20-second calibration on-off scans that use a local oscillator. 
	
Despite the limitations in target system tracking and observation duration due to its fixed dish, the 305-m William E. Gordon radio telescope at AO offered several distinct advantages relative to other facilities in 2010-2011.  The large effective area of the dish delivered exceptional sensitivity that yielded accurate polarization measurements across a large, $\sim$1 GHz simultaneous bandpass over short ($<$1 s) integration times.  Signals received by the dual-linear polarization C-band receiver\footnote{``Cband Calibration,'' available at https://naic.nrao.edu/arecibo/phil/calcb/cband.html\#top}, were processed by the recently commissioned Mock spectrometer\footnote{``Jeff Mock’s pdev Spectrometer,'' https://www.naic.edu/arecibo/phil/pdevall.html}.  The system temperature ranged within 25--32 K, with antenna gains of 5.5--9.0 K Jy$^{-1}$.  The half-power beam width varied with frequency from 0.97 $\times$ 1.09 arcmin (azimuth $\times$ zenith angle at 4.500 GHz central frequency) to 0.79 $\times$ 0.92 arcmin (5.400 GHz; Chris Salter, personal communication).

Full Stokes parameters were computed by the Mock spectrometer array, which consists of seven field-programmable gate array (FPGA) Fast Fourier Transform (FFT) Mock spectrometers individually tuned to central frequencies of 4.325, 4.466, 4.608, 4.750, 4.892, 5.034, and 5.176 GHz.  Each 8192-channel spectrometer yields a 172 MHz bandpass that overlaps the frequency range of each neighboring spectrometer by $\sim$30 MHz, thereby yielding continuous bandpass coverage from 4.239 to 5.262 GHz.  Prior to analysis, the signal-to-noise characteristics of the data were improved by rebinning the data from (frequency, time) resolutions of (20.9 kHz, 0.1 s) to (83.6 kHz, 0.9 s).  Radio frequency interference (RFI) was mitigated by an iterative statistical process detailed in \citet{rouphd}.

The analysis of each science scan consisted of the construction of dynamic spectra and bandpass-averaged time series graphs of flux density for every Stokes parameter for each spectrometer (Figure 1).  The per-scan sensitivity ranged from 1$\sigma\sim$0.1--1.2 mJy with sensitivity distribution reported in \citet{rou17b}.  Comprehensive, frequency and system-dependent sensitivity limits are available in \citet{rouphd}. These data products were then searched for bursts of $\gtrsim$10\% circularly polarized (Stokes V) emission, which are readily apparent when they occur and are a telltale sign of ECM emission.  Due to its local calibration procedure and confusion limitations, AO is insensitive to quiescent radio emission and radio bursts of duration comparable to the science scan duration.  Candidate bursts that exceeded a 3$\sigma$ noise threshold in their Stokes V time series graph then had their Stokes Q and U components examined for behavior suggestive of RFI, and their morphology compared with an archive of identified RFI artifacts observed across multiple systems.

\section{Results}
No magnetospheric radio emissions were detected from any system hosting exoplanet or brown dwarf companions.  I reiterate that this survey was only sensitive to $\gtrsim$10\% circularly polarized bursts of radio emission of several minutes duration or less, and was insensitive to quiescent (unpolarized or slowly varying) emission.  Nevertheless, using the standard deviation from the bandpass-averaged time series of the cleanest Mock spectrometer centered at 4.466 GHz, we can compute 3$\sigma$ detection limits on the circularly polarized radio emission from the substellar targets (Table 3).

In ROME III \citep{rou23}, I presented the results of my search for stellar radio emissions that may be caused by star-planet interactions in systems with hot exoplanet companions.  Each surveyed system resulted in a nondetection of the sought-after radio emission.  Since the host star and orbiting exoplanet(s) in each system fit within the $\sim$5 GHz half-power beamwidth, the results also constrain exoplanetary magnetospheric radio emissions.  Thus, Table 3 also recaps the magnetospheric detection limits for the ROME III targets (GJ 176 b, HD 46375 b, 55 Cnc b-f, GJ 436 b, HD 102195 b, HD 189733 b, HD 209458 b, and 51 Peg b), as well as HR 8799 b-e \citep{rou13}, so that they may be further contextualized, analyzed, and interpreted within the present work.

Figure 2 depicts the magnetospheric luminosity limits associated with the flux density detection limits in Table 3 as a function of spectral type for low-mass stars and brown dwarfs, and as a function of mass for exoplanets.  Since the objects in the present survey are bracketed on the high-mass end by UCDs, and at the low-mass end by Jupiter, I add the radio luminosities from these objects for contextualization of these results.  These upper limits provide tighter constraints on target radio emission than my later survey of UCD radio activity conducted with the same instrumental set up and analysis software due to the quieter radio environment around AO in 2010-2011 versus 2013-2014 \citep{rouphd,rou13,rou16b}.

\section{Discussion}

The survey frequency characteristics (Section 2.2) and 3$\sigma$ flux density detection limits (Table 3) may be compared with the frequency and flux density characteristics estimated for magnetospheric emissions from certain exoplanet targets (Table 4).  All references in Table 4 base their predicted exoplanet frequency and flux density properties on extrapolations of various scaling laws \citep{far99,laz04,rei10,gri17,zag18,kav19}.  Approximately one-third of targets were observed at frequencies within an order of magnitude of their anticipated highest frequency of magnetospheric ECM emission (HD 10697 b, HD 38529 c, HD 106252 b, HD 114762 Ab, 70 Vir b, HD 178911 Bb).  Two-thirds of targets were observed with enough sensitivity to detect their anticipated magnetospheric radio emissions, if they were observed at frequencies that correspond to magnetic field strengths present within their magnetospheres.  I note that many of these estimates predate theoretical work that suggests that ``inflated'' hot Jupiters may be explained by the existence of strong magnetic fields (e.g., \citealt{yad17}), and the discovery of powerful UCD magnetic fields which appear to violate these scaling laws (e.g., \citealt{kao18}).

\subsection{Targets Reclassified as Low-mass Stars or Brown Dwarfs}

Section 2.1 described how the measurement of $M~sin~i$ via the radial velocity method only determines the minimum companion mass, which can lead to an underestimate of its internal energy to power a convective dynamo.  This would then result in an underestimate of the maximum magnetic field strength of the substellar object, and accompanying radio surveys that probe frequencies lower than the cutoff frequency that corresponds to this maximum magnetic field strength.  While I did not detect magnetospheric radio emissions from any exoplanet candidate targeted by my survey, I highlight that several targets were subsequently reclassified as low-mass stars or brown dwarfs which would present a more optimistic case for the detection of $\sim$GHz-frequency radio emissions. 

\emph{HD 114762 Ab.--} Given the updated mass estimate for HD 114762 Ab from \citet{kie21} and the mass-spectral type relationship depicted in \cite[Fig.10]{chab00}, I estimate that this companion is actually an M5 dwarf, as opposed to a $\sim$11 M$_{J}$ giant planet as originally thought \citep{lat89}.  Although I did not detect any ECM-generated radio flares from this system, such flares would be common among M dwarfs.  For example, radio flares have been observed at 4.85 GHz from the M3.5 dwarf AD Leo \citep{ste01} and potentially from the M5 companion in the ``white dwarf pulsar'' cataclysmic variable system AR Sco at 1-10 GHz \citep{lyu20}.

\emph{HD 38529 c and HD 106252 b.--} Based on their radiometric Bode's law, \citet{laz04} estimated that if HD 38529 c and HD 106252 b were exoplanets, they would be sources of $\Phi_{radio}\sim$13 $\mu$Jy and $\Phi_{radio}\sim$32 $\mu$Jy magnetospheric radio emissions extending to cutoff frequencies of $\nu\sim$1.6 GHz and $\nu\sim$576 MHz, respectively.  They noted that given statistical variations in the empirical estimate of the magnetic moment for these objects, their actual cutoff frequencies could be within a factor of three higher or lower. However, the revised mass estimates for HD 38529 c and HD 106252 b (Table 1) clearly place them in the brown dwarf domain \citep{fri10,ref11}.  Using these revised masses together with 5 Gyr isochrone models \citep{bar03} and the effective temperatures from \citet{cus11}, I estimate that HD 38529 c and HD 106252 b have spectral types of $\sim$Y0 and $\sim$T9, respectively.  The 2010 observations of these two objects, therefore, constitute the first, although inadvertent, search for radio emission from potential Y dwarfs.  Characterizing the magnetism of this class of objects on the exoplanet-brown dwarf boundary may be important for several reasons. First, it may further illuminate the age, mass, and temperature ranges at which substellar object magnetic fields first begin to exceed those predicted by geodynamo scaling laws of rapid rotators, which may have important implications for their interior structures (e.g., \citealt{chr09,kao18}).  Second, the study of Y dwarfs in orbit around stars provide rare test cases of theories of brown dwarf and planet formation (e.g., \citealt{luh12}).  Third, radio observations may  constrain Y dwarf interior structure to better probe the tension between atmospheric theoretical models and retrievals (e.g., \citealt{leg21,wul24}). Since the coolest effective temperature and latest spectral type for which radio emission has been discovered is from the $T_{eff}\sim$900 K, T6.5 brown dwarf J1047+21 that produced $\Phi_{radio}\sim$3 mJy radio emissions at $\nu\gtrsim$4 GHz \citep{rou12}, these systems are plausible candidates for $\sim$GHz radio emissions at $\sim$mJy flux densities.  Unfortunately, both systems are $\sim$4$\times$ further away than J1047+21, so that while the instrumentation and techniques to search for radio emissions remain the same for all three objects, the radio emission detection limits for HD 38529 c and HD 106252 b are over an order-of-magnitude higher.

\subsection{Reasons for Nondetections}

There are several plausible explanations for the nondetection of magnetospheric radio emissions from the substellar objects in the target systems, including sensitivity, observing frequency, rotation phase coverage, and viewing geometry.  

\subsubsection{Sensitivity}
Starting with sensitivity, although this survey is the most sensitive to date, substellar target magnetospheric radio emissions would need to be $\sim$10$^{2}$ to 10$^{6}$ greater than those of Jupiter to be detectable (Table 3, Figure 2).  By comparison, \citet{zar18} estimated that hot Jupiters might produce radio emissions 10$^{3}$ to 10$^{6}$ times more intense than those emitted by Jupiter based on their empirically derived radio-magnetic Bode's law. Somewhat less optimistically, \citet{laz04} suggested that greater stellar wind loading among hot Jupiters would scale-up emissions analogous to Jovian peak power emissions by 10$^{2}$.  This survey, then, is at the edge of viability with respect to these scaling laws.

\subsubsection{Observing Frequency and Magnetic Field Strengths}
Another important facet to consider is the frequency coverage of this survey.  The survey was only sensitive to substellar magnetospheric emissions generated by the ECM mechanism.  The 4.239 GHz to 5.262 GHz frequency range of this survey corresponds to fundamental frequency (second harmonic) emission from magnetic fields of $B\sim$1.5 to 1.9 kG ($B\sim$760 to 940 G) according to the cyclotron formula in Section 2.1.  Theoretical models indicate that both modes are possible, although which mode of emission dominates is dependent upon local plasma conditions \citep{tre06}.  Indeed, \citet{tri11} assert that both emission modes operate near-simultaneously in the magnetosphere of the chemically peculiar star CU Virginis.

Given prior work on empirical scaling laws and interior dynamo models, it is not surprising that most surveys of exoplanet magnetospheric emissions were conducted at lower frequencies.  Only a handful of targets were hypothesized to have cutoff frequencies extending into the $\sim$GHz range such as HD 114762 Ab, HD 38529 c (before their updated mass determinations), HD 43197 b, PSR B1620-26 b, and WASP-77A b \citep{laz04,sir14}.  However, given the $M~sin~i$ mass ambiguity, if, for example, the mass of an object such as HD 38529 c was underestimated by a factor of 1.8, this would correspond to an increase in its emission frequency of $\sim$2.7, placing it within the range of this survey.  Indeed, after the completion of the survey it was confirmed that HD 38529 c, HD 106252 b, and HD 114762 Ab were more massive than first thought, with the first two targets acknowledged to be brown dwarfs and the last I estimate to be an M5 dwarf.  In addition, a comparison of the dipole component of the Jovian surface magnetic field ($B\sim$4.18 G) with its small-scale polar surface field ($B\sim$20 G) indicates that small-scale fields on exoplanets may result in significantly higher ECM emission frequencies \citep{con22}.

Similarly, confirmation that brown dwarfs with exoplanet-like effective temperatures or masses, such as J1047+21 and SIMP 01365662+0933473 maintain $\sim$3 to 6 kG magnetic fields may imply that giant exoplanets maintain magnetic fields much stronger than anticipated, that may be used to constrain and validate magnetic dynamo models (e.g., \citealt{chr09,rei10}).  I reserve a more thorough discussion of dynamo scaling laws as applied to exoplanets and brown dwarfs for a later paper in the ROME series.  However, these considerations indicate that it is not unreasonable to survey exoplanets for magnetospheric emissions at higher frequencies than traditionally assumed, given enough sensitivity.  On the other hand, since the dynamo models of \citet{chr09} and \citet{rei10} require rapid rotation, and hot Jupiters may be tidally locked, the convective motions of their internal dynamos may be drastically reduced, resulting in diminished global dipolar magnetic fields (e.g., \citealt{san04}).  Alternatively, \citet{yad17} argued that the surprisingly large radii of hot Jupiters indicate that stellar deposition of heat into their interiors should enhance convective dynamo activity, leading to inflated interiors and stronger magnetic fields. It is also possible that the dense stellar wind environment in which these hot Jupiters orbit their host stars may preclude the escape of any ECM radio emissions (e.g., \citealt{web17,kav19}).  Thus, the search for magnetospheric radio emissions from substellar objects at a range of observable parameters will provide valuable information about myriad system properties.

However, since the survey failed to detect any magnetospheric radio emissions, it is likely that most of the substellar targets, with the exceptions of HD 114762 Ab, HD 38529c, HD 196252b, and HR 8799 b, c, d, and e, were observed at frequencies in excess of their cutoff frequencies, in addition to inadequate sensitivity.

\subsubsection{Rotation Phase Coverage}
ECM-generated magnetospheric radio emission is beamed into a hollow cone at large opening angles relative to the magnetic field, modulated at the rotation period of the object \citep{zar98}.  Exoplanet rotation periods range from $\sim$10 hrs for Jupiter and Saturn to 3--4 d for spin-orbit synchronized, sidereal rotation of hot Jupiters \citep{san04,dep15}.  As described in Table 2, every target system was observed for 0.67 to 2.5 hrs. Thus, in the best case scenario of a distant giant exoplanet that has not undergone significant orbital evolution, the observations only cover $\sim$10-20\% of rotational phase for potential magnetic activity (e.g., $\epsilon$ Tau).  

The rotational phase coverage of the  brown dwarf targets HD 38529 c and HD 106252 b is probably significantly better.  Magnetically-active UCD rotation periods range from 0.288 hrs (J1122+25) to $\sim$50 hrs (SDSS J151643.01+305344.4), with a mean rotation period $P\sim$1.22 hr \citep{rou17a,rou17b}.  Since HD 38529 c and HD 106252 b were both observed for 1 hr, if they rotate at the mean UCD rotation period, then their rotational phase coverage is likely $\sim$80\%, although given the spread in rotation periods, it could be as low as $\sim$2\%. A similar result applies to the $\sim$M5 HD 114762 Ab.  Thus, the duration of the observations provide rotational phase coverage of $\sim$2 to 80\% for exoplanet or brown dwarf magnetospheric radio emissions.

\subsubsection{Viewing Geometry}
Based on an examination of the solar system planets, we may assume that the magnetic and rotation axes of exoplanet and brown dwarf companions are roughly aligned with each other, and with their orbital axes in low-obliquity orbits.  Since ECM-induced magnetospheric emissions are beamed into a hollow cone at large half-apex angles with respect to the magnetic axis of the exoplanet or brown dwarf, these emissions are optimally viewed at high inclination to the orbital axis (e.g., \citealt{zar98}).  This suggests that transiting and near-transiting systems, such as HD 189733 and the other targets in Table 4, are optimal targets to search for auroral radio emissions.  On the other hand, directly-imaged systems, such as HR 8799 with its face-on geometry as observed from Earth, may have detectable radio emissions only if the exoplanet magnetic, rotation, and/or orbital axes are significantly misaligned such that their magnetic axes do not point toward the Earth.  Therefore, while viewing geometry may hinder the detectability of radio emissions on a case-by-case basis, it should not adversely affect the detectability of the target population as a whole.  However, it may be a contributing factor in the nondetection of emissions from systems viewed at low inclination angles with respect to their companions' orbital axes, such as HR 8799 b-e.

\subsubsection{Detection Efforts Past, Present, and Future}
Numerous attempts have been made to detect the magnetospheric emissions from exoplanets at radio wavelengths since the work of \citet{yan77}.  Although it is beyond the scope of this paper to evaluate the merits of these efforts, I will briefly compare my instrumentation and strategy to those of previous efforts, then discuss how this survey, together with surveys of UCDs, may guide future detection efforts.

This survey at AO has several distinguishing features relative to previous surveys.  AO provided exquisite sensitivity at short timescales by achieving 3$\sigma$ detection limits of $\sim$1 mJy in 0.9 s integrations.  This compares very favorably with many previous large surveys that achieved detection limits of 3$\sigma\sim$ 10 to 100 mJy in $\sim$1 ks integrations at, for instance, the Murchison Wide Array (MWA, \citealt{lyn18}), the Ukrainian T-shaped Radio Telescope second modification (UTR-2), and the Very Large Array (VLA; \citealt{zar15} and references therein).  Targets in this survey were simultaneously observed over a $\Delta\nu\sim$1 GHz bandpass, as opposed to many earlier efforts that were limited to $\Delta\nu\leq$50 MHz due to technology at the time.  Finally, the spectropolarimetric survey yielded dynamic spectra in all four Stokes parameters for every target, while  earlier surveys often measured total intensity (Stokes I) only.  The survey focused on circularly polarized (Stokes V) emission and used linear polarization (Stokes Q and U) as an RFI mitigation strategy.  Newer facilities with better technology have continued to improve search characteristics in the same general directions as this survey (e.g., \citealt{tur21}).

\citet{rou16b} discussed trends in $\nu~L_{\nu}$ among UCDs, which display similarities in their magnetism as manifest by X-ray, H$\alpha$, and radio activity, despite differences in their near infrared and optical spectra (e.g., \citealt{mcl12}).  My detection and characterization of flaring radio emission from two late T brown dwarfs, J1047+21 and J1122+25, may indicate how trends in substellar radio activity evolve toward exoplanet-mass objects.  Figure 2 updates the plot of detected radio emissions from objects near the stellar-substellar boundary and includes a side-by-side comparison of my present exoplanet and previous UCD survey efforts \citep[Fig. 3]{rou16b}.

Jupiter's DAM is 4-5 orders of magnitude below the UCD flaring $\nu~L_{\nu}$ radio emission trend line in Figure 2. \citet{rou16b} proposed that radio emission ($\nu~L_{\nu}$) may remain approximately constant from M7 to T7, or perhaps declines from L2 to T3, followed by a minor rise from T3 to $\sim$T6.  Yet at some spectral type $\gtrsim$T7, flaring and quiescent radio emission from UCDs and giant exoplanets should be greatly reduced so that the trend at $\sim$T6 smoothly connects with Jovian emissions.  Unfortunately, the nondetection of the brown dwarfs HD 106252 b ($\sim$T9) and HD 38529 c ($\sim$Y0) does not shed any light on this matter as their radio detection limits exceed the UCD trend line and the emissions from J1047+21 and J1122+25 (Figure 2). My detection limits for even the most massive exoplanets surveyed, such as those of the HR 8799 system and 70 Vir b, are likely $\gtrsim$3 orders of magnitude above the hypothetical trend line in Figure 2.  This estimate is consistent with the range suggested by \citet{zar18} as well as that of  \citet{laz04} once statistical uncertainties are accounted for.

\subsection{Implications for Recent Theoretical Work}
Recently, \citet{ash22} sought to improve upon previous theoretical efforts to compute the maximum fundamental frequencies and flux densities of ECM emissions from 671 exoplanets.  This effort combined scaling laws of exoplanet magnetic properties (i.e., convective zone sizes, magnetic moments, magnetopause standoff distances, surface magnetic fields, and maximum frequency of ECM emission) with GAMERA 3D magnetohydrodynamic simulations of stellar wind activity to produce potentially ultra-precise radio spectra of targets.  Yet observational work that seeks to verify these predictions must proceed with care. The computed exoplanet surface magnetic fields are scaled to that of Jupiter, reported to be 4.18 G \citep{con22}.  From this value, maximum frequencies of exoplanet ECM emissions are derived from a mass-radius-rotation scaling relation anchored on the professed maximum frequency of Jovian ECM emissions, 24 MHz.  This value is approximately half the observationally determined cutoff frequency (Section 1).  It might be somewhat unexpected to use a scaling relation to determine the maximum fundamental frequency of emissions, $\nu_{peak}$, when it can be directly computed from the surface magnetic field via the cyclotron formula (Section 2.1).  However, it is noteworthy that the maximum surface magnetic field of Jupiter is $\sim$20 G in the polar regions \citep[Fig. 5]{con22}, which leads to a modeled underestimate of the maximum exoplanet surface magnetic field strengths ($B_{S}$) by a factor of $\sim$5.  Although the reasoning presented in \citet[Appendix B]{ash22} directly follows prior work by \citet{laz04}, the omission of the earlier work's disclaimer that the maximum frequency of emission may vary by a factor of three due to statistical uncertainties in their empirical relation can underestimate uncertainty in modern models.  
		
\citet{ash22} also omit other important factors in the detection of substellar magnetospheric radio emissions.  In a number of systems, unless the exoplanet ECM frequency is greater than the local plasma frequency, no ECM signal will be detected.  For example, \citet{kav19} determined that the exoplanet in the HD 189733 system requires $\nu_{peak}>$21 MHz for the its magnetospheric emissions to escape the stellar wind plasma frequency.  Exoplanet radio emissions with $\nu_{peak}\lesssim$10 MHz would be reflected by the terrestrial ionosphere \citep{dep15} and hence, undetectable by low-frequency ground-based surveys (e.g., 51 Peg, AU Mic b in their Table 5, \citealt{ash22}).  Both phenomena are not accounted for in \citet{ash22}.  

Finally, we note that several case studies presented in \citet{ash22} require additional attention. Their Table 2 estimates the maximum magnetic field strengths of, for example, brown dwarfs of masses 45--62 M$_{J}$ such as NGTS-7A b, SCR 1845 b, and ZTF J0038+2030 b, as ranging from 28.6 to 80.2 G.  The discovery of nearly two dozen brown dwarfs with $\sim$kG magnetic fields indicates that their model likely underestimates these magnetic field strengths.  \citet{ash22} also compute the flux density and peak frequency of emission for $\tau$ Bo\"{o}tis b, to demonstrate that the model is consistent with the purported detection of ECM emission from the exoplanet \citep{tur21}. However, should $\tau$ Boo b have a surface magnetic field of $B_{S}\sim$ 8 G, the scaling relation in \citet[Eq. B15]{ash22} would yield $\nu_{peak}$=46 MHz, which is in tension with the result reported in their Tables 2 and 3.  Given these discrepancies in the results reported in Tables 2, 3, and 5, it is unclear whether the modeled frequencies and flux densities should have been detectable during the 18-32 MHz UTR-2 survey conducted by \citet{rya04}, although no detection of the target was reported.

\subsection{Implications for the Direct Detection of Exomoons within Habitable Zones}

Two targeted systems host exoplanets that potentially host exomoons within their host stars' HZs for the entire duration of their orbits: HD 10697 b and 55 Cnc f.  Several others spend a fraction of their orbital periods within the optimistic HZ, such as HD 38529 c (57\%), HD 50554 b (26\%), and HD 106252 b (20\%)\footnote{``Habitable Zone Gallery,'' available at http://www.hzgallery.org/table.html} \citep{kan12,kop14}.  Both HD 10697 b and 55 Cnc f are significantly more massive than the Earth, so that any life within those systems likely resides on orbiting exomoons.  Perhaps such exomoons could generate powerful radio emissions akin to the Io-controlled DAM, and their presence could be inferred through the modulation of exoplanetary magnetospheric radio emissions by the exomoons' orbital periods.  However, these radio emissions would likely still follow the established radio-kinetic and radio-magnetic scaling laws that relate the overall power of their auroral radio emissions to the kinetic or magnetic energy flux received by their magnetospheric cross sections as they traverse the exoplanet's magnetosphere.  In the case of Io and Ganymede orbiting Jupiter, their auroral radio emissions do not exceed that of Jupiter \citep{zarplus18}.  Hence, the exomoons in such systems would modulate, but offer no power enhancement, of the radio emissions of these exoplanets.

\section{Conclusion}

In 2010-2011, I surveyed exoplanets and brown dwarfs in a quest to detect and characterize their magnetospheric radio emissions, and thereby gain insight into their magnetic activity, magnetic fields, surrounding plasma environments, and interior structures.  In this fourth installment in the Radio Observations of Magnetized Exoplanets (ROME) series, I present further results on purported exoplanet-hosting systems, which started with HD 189733 A/B/b (ROME I/II) and continued with a search for star-planet interactions (ROME III).  The present survey sought Jupiter-like auroral ECM radio emissions derived from magnetospheric interactions between stellar winds and substellar companion magnetospheres, a type of plasma flow-obstacle interaction.

This survey targeted nine systems with substellar companions, with a preference for massive companions with semimajor axes $a\lesssim$0.1 au.  Massive exoplanets are hypothesized to hold larger reservoirs of internal energy to power convective dynamos that generate stronger magnetic fields and larger magnetospheric cross sections for plasma flow-obstacle interactions than generally has been anticipated.  Substellar companions close to their host stars would encounter greater kinetic flow power and Poynting fluxes for magnetospheric interactions.

This survey leveraged the exquisitely sensitive 305 m Arecibo radio telescope tuned to 4.2-5.2 GHz to search for strong, $\sim$kG magnetic fields, similar to those found among the coolest brown dwarfs, such as J1047+21 and J1122+25.  Dynamic spectra in all four Stokes parameters from the Mock spectrometers were used to search for $\gtrsim10$\% circularly polarized ECM emission and to mitigate RFI.  All targets were observed for $\lesssim$2.5 hr, the maximum time for them to transit the fixed dish.

Several of the surveyed systems are now known to host low-mass star or brown dwarf companions: HD 114762 Ab ($\sim$M5), HD 106252 b ($\sim$T9), and HD 38529 c ($\sim$Y0).  Hence, this is the first survey to search for radio emission from ultracool dwarfs of spectral type Y, which may illuminate unanswered questions regarding their magnetism, interior and atmospheric structure, and formation histories.  Although similar objects (e.g., J1047+21) are powerful radio emitters at $\sim$5 GHz, the lack of detected emissions in these three targets may stem from magnetic inactivity, sporadic emissions, or lack of sensitivity.  Additional exoplanet targets were anticipated to be especially promising for the generation of high-frequency radio emissions based on empirically-derived radiometric Bode's laws, including 70 Vir b, HD 178911 Bb, and HD 10697 b.  However, no exoplanet magnetospheric radio emissions were detected.  The 3$\sigma$ sensitivity for targeted systems ranged from 1.05 to 1.43 mJy, resulting in $\nu~L_{\nu}$ upper limits of 1.8$\times$10$^{24}$ erg s$^{-1}$ to 1.4$\times$10$^{25}$ erg s$^{-1}$.

This search may not have succeeded for several reasons.  First, construction of a trend line connecting the radio activity from the coolest, latest type brown dwarfs to that of Jupiter indicates that instrumental sensitivity must increase by $\gtrsim$10$^{3}$ in order to detect magnetospheric radio emissions.  Second, although the exoplanet targets may have been surveyed at too-high of frequencies, it remains unknown what are the ECM cutoff frequencies of giant exoplanets more massive than Jupiter.  Third, since giant exoplanet rotation periods are likely $\sim$10 hrs, the rotational phase coverage of their magnetospheric activity varied from 2--80\%.

The analysis of the survey results highlights how the productivity of future searches for magnetospheric exoplanet emissions may be improved.  First, increases in sensitivity of $\sim10^{2}$ to $10^{6}$ are required to detect magnetospheric emissions.  Second, future surveys should simultaneously observe as wide of a bandpass as possible given the great uncertainty in exoplanet magnetic field strengths, and therefore, the frequency of their ECM emissions.  Third, future surveys should leverage full Stokes polarimetry to distinguish among and characterize various types of astrophysical emission, and as an RFI mitigation strategy.  Fourth, multiepoch, high-cadence observations are required to untangle the effects of intrinsic stellar activity from variable magnetospheric exoplanet radio emissions, since stars are typically assumed to be less dynamic than they actually are.
 
\section{Acknowledgments}

M.R. would like to acknowledge support from the Center for Exoplanets and Habitable Worlds and the Zaccheus Daniel Fellowship. The Center for Exoplanets and Habitable Worlds is supported by Pennsylvania State University and the Eberly College of Science.  Data storage and analysis support has been made possible with the Theodore Dunham, Jr. Grants for Research in Astronomy.  At the time of the observations that are the subject of this publication, the Arecibo Observatory was operated by SRI International under a cooperative agreement with the National Science Foundation (AST-1100968), and in alliance with Ana G. M\'{e}ndez-Universidad Metropolitana, and the Universities Space Research Association.

This research has made use of NASA's Astrophysics Data System and the SIMBAD database, operated at CDS, Strasbourg, France.  Guidance on properties and publications related to objects of interest were obtained from the catalog located at exoplanet.eu.

\facility{Arecibo}
\software{IDL}

\clearpage

\begin{deluxetable}{lllllllll}
\tabletypesize{\scriptsize}
\tablecolumns{9}
\tablewidth{0pt}
\tablecaption{Survey Target Properties}
\tablehead{
	\colhead{Name}&
	\colhead{R.A.}&
	\colhead{Dec.}&
	\colhead{Host Star} &
	\colhead{Dist.}&
	\colhead{Semimajor}&
	\colhead{Period}&
	\colhead{Mass}&
	\colhead{Properties}\\
	\colhead{}&
	\colhead{({hh}\phn{mm}\phn{ss})}&
	\colhead{(\phn{\arcdeg}~\phn{\arcmin}~\phn{\arcsec})}&
	\colhead{Type}&
	\colhead{(pc)}&
	\colhead{Axis (AU)}&
	\colhead{(d)}&
	\colhead{(M$_{J}$)}&
	\colhead{Refs}
}
\startdata
HD 10697 b\tablenotemark{a} & 01 44 55 & +20 04 59 & G5 IV & 33.16 & 2.12 & 1072.3 & 6.837 & \underline{{\bf {\em 1}}},2\\
$\epsilon$ Tau b\tablenotemark{b} & 04 28 37 & +19 10 50 & K0 III & 44.71 & 1.90 & 594.9 & 7.34 & {\bf {\em 3}},\underline{4},4\\
HD 38529 b & 05 46 35 & +01 10 06 & G4 IV & 42.42 & 0.131 & 14.31 & 0.90 & {\bf 5},\underline{{\em 6}},6\\
$\phantom{HD 38529x}$c & $\phantom{05 46 35}$ & $\phantom{+01 10 06}$ & $\phantom{G4 IV}$ & $\phantom{42.42}$ & 3.697 & 1236.14 & 17.6\tablenotemark{c} & {\bf 7},\underline{{\em 6}},6\\
HD 50554 b & 06 54 43 & +24 14 44 & F8 V & 31.07 & 2.28 & 1224 & 4.46 & {\bf 9},\underline{{\em 8}},8\\
HD 106252 b & 12 13 30 & +10 02 30 & G0 V & 38.10 & 2.6 & 1531 & 30.6\tablenotemark{d} & {\bf 9},\underline{8},{\em 10},10\\
HD 114762 Ab & 13 12 20 & +17 31 02 & F9 V & 38.17 & 0.361 & 83.915 & 147\tablenotemark{e} & {\bf 11},\underline{\em 12},12\\
70 Vir b & 13 28 26 & +13 46 44 & G4 V & 18.10 & 0.481 & 116.69 & 7.4 & {\bf 13},\underline{\em 14},14\\
HD 178911 Bb & 19 09 03 & +34 35 59 & G5 & 40.96 & 0.339 & 71.484 & 7.03 & {\bf 15},\underline{{\em 16}},16\\
HD 195019 b & 20 28 19 & +18 46 10 & G3 IV/V & 37.53 & 0.1388 & 18.2 & 3.69 & {\bf 17},\underline{{\em 8}},8\\
\enddata
\tablecomments{Exoplanet, brown dwarf, and low-mass stellar companion hosting system properties.  In the properties references column (rightmost column), {\bf bold}, \underline{underlined}, and \emph{italicized} numerals denote discovery, semimajor axis, and orbital period references, respectively. The final number in the column in normal font provides the companion object's mass reference.  All distances are from the \citet{gai21}.  {\bf References.} (1) \citet{vog00}; (2) \citet{sim10}; (3) \citet{sat07}; (4) \citet{kun11}; (5) \citet{fis01}; (6) \citet{fri10}; (7) \citet{fis03}; (8) \citet{but06}; (9) \citet{per03}; (10) \citet{ref11}; (11) \citet{lat89}; (12) \citet{kie21}; (13) \citet{mar96}; (14) \citet{kan15}; (15) \citet{zuc02}; (16) \citet{wit09}; (17) \citet{fis99}}
\tablenotetext{a}{Also known as 109 Psc b.}
\tablenotetext{b}{Also known as HD 28305 b.}
\tablenotetext{c}{With a mass of 0.017$M_{\odot}$, 5 Gyr isochrones yields $T_{eff}\sim$430 K and spectral type $\sim$Y0 \citep{bar03,cus11}.}
\tablenotetext{d}{With a mass of 0.029$M_{\odot}$, 5 Gyr isochrones yields $T_{eff}\sim$610 K and spectral type $\sim$T9 \citep{bar03,cus11}.}
\tablenotetext{e}{\citet[Fig. 10]{chab00} indicates spectral type $\sim$M5 corresponds to a 0.140$M_{\odot}$,  $\sim$5 Gyr old object of solar metallicity.}
\end{deluxetable}

\begin{deluxetable}{lccccc}
\tabletypesize{\scriptsize}
\tablecolumns{6}
\tablewidth{0pt}
\tablecaption{Observations List}
\tablehead{
	\colhead{Name}&
	\colhead{UT Date}&
	\colhead{File Date}&
	\colhead{Scan Range\tablenotemark{a}}&
	\colhead{Number} &
	\colhead{Time on}\\
	\colhead{}&
	\colhead{(yyyymmdd)}&
	\colhead{(yyyymmdd)}&
	\colhead{}&
	\colhead{of Scans}&
	\colhead{Source (hrs)}
}
\startdata
HD 10697 & 2010 12 21 & 2010 12 20 & 00000-01200 & 4 & 0.67\\
$\epsilon$ Tau & 2010 01 09 & 2010 01 08 & 00000-02000 & 7 & 1.17\\
& 2010 12 18 & 2010 12 17 & 10100-10700{*} & 3 & 0.50\\
& 2010 12 18 & 2010 12 18 & 00000-00400 & 1 & 0.17\\
& 2010 12 21 & 2010 12 20 & 02500-03600 & 4 & 0.67\\
HD 38529 & 2010 01 09 & 2010 01 08 & 02100-03800 & 6 & 1.00\\
HD 50554 & 2010 01 09 & 2010 01 08 & 03900-04200{*} & 2 & 0.33\\
& 2010 01 09 & 2010 01 09 & 00000-00400 & 1 & 0.17\\
& 2010 12 18 & 2010 12 18 & 00500-02800 & 8 & 1.33\\
& 2010 12 19 & 2010 12 19 & 00800-01900 & 4 & 0.67\\
HD 106252 & 2010 01 07 & 2010 01 07 & 01800-03500 & 6 & 1.00\\
HD 114762 & 2010 01 07 & 2010 01 07 & 03600-03800 & 1 & 0.17\\
& 2011 01 02 & 2011 01 02 & 08700-10500 & 7 & 1.17\\
70 Vir & 2010 01 06 & 2010 01 06 & 03600-05600 & 7 & 1.17\\
HD 178911 B & 2011 07 20 & 2011 07 19 & 06200-07100{*} & 4 & 0.67\\
& 2011 07 20 & 2011 07 20 & 00000-01000 & 3 & 0.50\\
HD 195019 & 2011 07 20 & 2011 07 20 & 01100-04000 & 10 & 1.67\\
\enddata
\tablecomments{Characteristics of AO data sets acquired and analyzed to search for magnetospheric radio emissions.}
\tablenotetext{a}{The scan range column denotes the continuous sequence of on-source, calibration-on, and calibration-off scans that focused on the listed target on a given day.  For example, file a2471.20100107.b0s1g0.03600.fits provides the on-source data from the first, lowest-frequency Mock spectrometer (b0) during the HD 114762 Ab observation (scan 03600) that occurred on 2010 Jan 7.  Asterisks in the scan range column denote an observing session that immediately continued onto the following day.}
\end{deluxetable}

\begin{deluxetable}{llllll}
\tabletypesize{\scriptsize}
\tablecolumns{8}
\tablewidth{0pt}
\tablecaption{Survey Detection Results}
\tablehead{
	\colhead{Object}&
	\colhead{Mass}&
	\colhead{Flux Density \tablenotemark{a}}&
	\colhead{$\nu$L$_{\nu}$ \tablenotemark{a}}&
	\colhead{$\nu$L$_{\nu}$} \tablenotemark{b}&
	\colhead{$\nu$L$_{\nu}$\ \tablenotemark{b,c}}\\
	\colhead{}&
	\colhead{(M$_{J}$)}&
	\colhead{Limit (mJy)}&
	\colhead{(ergs s$^{-1}$)}&
	\colhead{(log L$_{\odot}$)}&
	\colhead{(log L$_{J,rad}$)}
}
\startdata
HD 10697 b & 6.837 & $<$1.305 & $<$7.67$\times$10$^{24}$ & $<$-8.70 & $<$6.97\\
$\epsilon$ Tau b & 7.34 & $<$1.101 & $<$1.18$\times$10$^{25}$ & $<$-8.51 & $<$7.16\\
GJ 176 b\tablenotemark{d} & 0.0264 & $<$1.065 & $<$5.13$\times$10$^{23}$ & $<$-9.87 & $<$5.80\\
HD 38529 b & 0.90 & $<$1.425 & $<$1.37$\times$10$^{25}$ & $<$-8.45 & $<$7.22\\
$\phantom{HD 38529x}$c & 17.6 & $<$1.425 & $<$1.37$\times$10$^{25}$ & $<$-8.45 & $<$7.22\\
HD 46375 b\tablenotemark{d} & 0.226 & $<$1.134 & $<$5.28$\times$10$^{24}$ & $<$-8.86 & $<$6.81\\
HD 50554 b & 4.46 & $<$1.038 & $<$5.36$\times$10$^{24}$ & $<$-8.86 & $<$6.82\\
55 Cnc e\tablenotemark{d} & 0.027 & $<$0.984 & $<$8.34$\times$10$^{23}$ & $<$-9.66 & $<$6.01\\
$\phantom{55 Cncx}$b & 0.8036 & $<$0.984 & $<$8.34$\times$10$^{23}$ & $<$-9.66 & $<$6.01\\
$\phantom{55 Cncx}$c & 0.1611 & $<$0.984 & $<$8.34$\times$10$^{23}$ & $<$-9.66 & $<$6.01\\
$\phantom{55 Cncx}$f & 0.1503 & $<$0.984 & $<$8.34$\times$10$^{23}$ & $<$-9.66 & $<$6.01\\
$\phantom{55 Cncx}$d & 3.12 & $<$0.984 & $<$8.34$\times$10$^{23}$ & $<$-9.66 & $<$6.01\\
GJ 436 b\tablenotemark{d} & 0.0737 & $<$1.047 & $<$5.35$\times$10$^{23}$ & $<$-9.86 & $<$5.82\\
HD 102195 b\tablenotemark{d} & 0.46 & $<$1.302 & $<$6.00$\times$10$^{24}$ & $<$-8.81 & $<$6.86\\
HD 106252 b & 30.6 & $<$1.194 & $<$9.26$\times$10$^{24}$ & $<$-8.62 & $<$7.05\\
HD 114762 Ab & 147 & $<$1.140 & $<$8.88$\times$10$^{24}$ & $<$-8.64 & $<$7.03\\
70 Vir b & 7.4 & $<$1.047 & $<$1.83$\times$10$^{24}$ & $<$-9.32 & $<$6.35\\
HD 178911 Bb & 7.03 & $<$1.287 & $<$1.15$\times$10$^{25}$ & $<$-8.52 & $<$7.15\\
HD 189733 b\tablenotemark{d} & 1.13 & $<$1.158 & $<$2.42$\times$10$^{24}$ & $<$-9.20 & $<$6.47\\
HD 195019 b & 3.69 & $<$1.203 & $<$9.06$\times$10$^{24}$ & $<$-8.63 & $<$7.04\\
HD 209458 b\tablenotemark{d} & 0.714 & $<$1.155 & $<$1.43$\times$10$^{25}$ & $<$-8.430 & $<$7.24\\
51 Peg b\tablenotemark{d} & 0.46 & $<$2.067 & $<$2.66$\times$10$^{24}$ & $<$-9.16 & $<$6.51\\
\hline
HR 8799\tablenotemark{e} e & 9.6 & $<$1.044 & $<$9.33$\times$10$^{24}$ & $<$-8.62 & $<$7.06\\
$\phantom{HR 8799x}$d & 7.2 & $<$1.044 & $<$9.33$\times$10$^{24}$ & $<$-8.62 & $<$7.06\\
$\phantom{HR 8799x}$c & 7.2 & $<$1.044 & $<$9.33$\times$10$^{24}$ & $<$-8.62 & $<$7.06\\
$\phantom{HR 8799x}$b & 5.8 & $<$1.044 & $<$9.33$\times$10$^{24}$ & $<$-8.62 & $<$7.06\\
\enddata
\tablecomments{Detection limits and radio luminosities of target systems.}
\tablenotetext{a}{Since 3$\sigma$ detection limits are reported, the uncertainty in the flux density limits is plus/minus one-third the reported value.}
\tablenotetext{b}{The uncertainty in each scaled luminosity value is +0.13/-0.18.}
\tablenotetext{c}{Luminosity in terms of the average power output of Jupiter's ECM-induced DAM during maximum solar activity (as opposed to average power or peak power), or $L_{J,rad}=8.2\times$10$^{17}$ erg~s$^{-1}$ \citep{zar04}.}
\tablenotetext{d}{Radio emission flux density detection limits presented in \citet{rou23}.}
\tablenotetext{e}{HR 8799 exoplanet radio emission flux density detection limits were reported during my first UCD survey \citep{rou13}.}
\end{deluxetable}
\clearpage

\begin{deluxetable}{llll}
	\tabletypesize{\scriptsize}
	\tablecolumns{4}
	\tablewidth{0pt}
	\tablecaption{Anticipated Exoplanet Radio Emission Properties from the Literature}
	\tablehead{
		\colhead{Object}&
		\colhead{Maximum}&
		\colhead{Maximum Flux}&
		\colhead{References}\\
		\colhead{}&
		\colhead{Frequency (MHz)}&
		\colhead{Density (mJy)}&
		\colhead{}
	}
	\startdata
	HD 10697 b & 545,88.4 & 0.08 & 2,4\\
	$\epsilon$ Tau b & 91.7 & 0 & 4\\
	GJ 176 b & 1.3 & 26.3 & 4\\
	HD 38529 b & 16,20.4 & 6,0.5 & 2,4\\
	$\phantom{HD 38529x}$c & 1600,209.1 & 0.013,0 & 2,4\\
	HD 46375 b & 0.6,0.8 & 80,178.2 & 2,4 \\
	HD 50554 b & 333,73.5 & 0.063,0 & 2,4\\
	55 Cnc e & 29.6 & 148.2 & 4\\
	$\phantom{55 Cncx}$b & 0.5,17.6,19.6 & 1.5,80,2.9 & 1,2,4\\
	$\phantom{55 Cncx}$c & 1.7,6.7 & 40,0 & 2,4\\
	$\phantom{55 Cncx}$f & 6.8 & 0 & 4\\
	$\phantom{55 Cncx}$d & 242,70.2 & 0.08,0 & 2,4\\
	GJ 436 b & 18.2 & 783.2 & 4\\
	HD 102195 b & 11.4 & 36.6 & 4\\
	HD 106252 b & 576,265.7 & 0.032,0 & 2,4\\
	HD 114762 Ab & 1090,121.4 & 0.015,0 & 2,4\\
	70 Vir b & 545,546,86.1 & 0.023,2,0 & 1,2,4\\
	HD 178911 Bb & 504,82.9 & 0.63,0 & 2,4\\
	HD 189733 b & 39,21.4,15,25 & 57,520.3,20,50-130 & 3,4,5,6\\
	HD 195019 b & 184,57.3 & 16,0.1 & 2,4\\
	HD 209458 b & 2.9,2.9 & 25, 49.4 & 2,4\\
	51 Peg b & 0.7,1.2,7 & 3.9,251,0.8 & 1,2,3\\
	\enddata
	\tablecomments{Estimated maximum fundamental frequencies and flux densities of ECM radio emission as reported in the literature.  For objects with multiple estimates, the listed frequencies and flux densities are given in the corresponding order to the references. For references that provide multiple estimates that depend on, for instance, radiation model \citep{far99} or planetary rotation \citep{gri17}, only the maximum frequency/flux density combination is reported. Signals with $\nu_{c}\lesssim$10 MHz are reflected by the terrestrial ionosphere, and hence would be undetectable from the ground. Note that even though only a few estimates are listed, the maximum frequency of emission can vary by an order of magnitude, and the maximum flux density can vary from a target having detectable to no detectable emission.\\
		{\bf References.} (1) \citet{far99}; (2) \citet{laz04}; (3) \citet{rei10}; (4) \citet{gri17}; (5) \citet{zag18}; (6) \citet{kav19}.}
\end{deluxetable}
\clearpage

\begin{figure}
	\centering
	\includegraphics[trim = 00mm 0mm 0mm 0mm, clip, width=0.6\textwidth,angle=0]{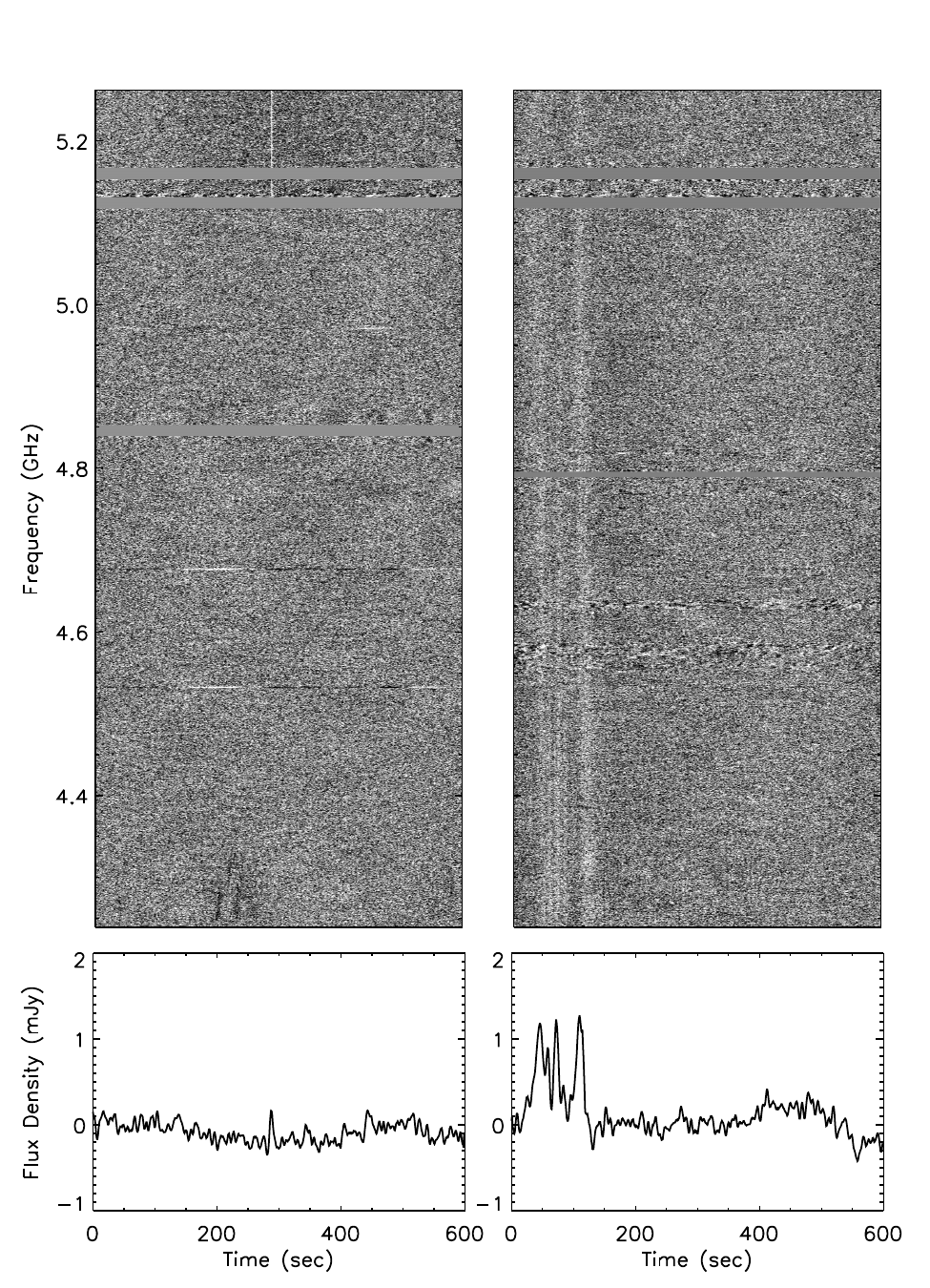}
	\caption{Examples of Stokes V dynamic spectra and bandpass-integrated time series from this survey (left) and a previous survey of UCDs (right) at AO.  Horizontal gray bars near 5.1 to 5.2 GHz in both spectra, 4.85 GHz at left, and 4.8 GHz at right represent the excision of strong RFI.  Even though a patch of left-circularly polarized emission occurs at $t\sim$200 s in the dynamic spectrum from $\epsilon$ Tau at left, the strong linear polarization of this feature coupled with its occurrence in other data sets (e.g., for GJ 176 in \citealt[Fig.1]{rou23}) indicate that it is an RFI artifact. An ECM-induced $\sim$70\% circularly polarized radio flare from the T6.5 UCD J1047+21 is depicted at right for comparison \citep{rou12}.}
\end{figure}

\begin{figure}
	\centering
	\includegraphics[trim = 1mm 3mm 1mm 3mm, clip, width=0.9\textwidth,angle=0]{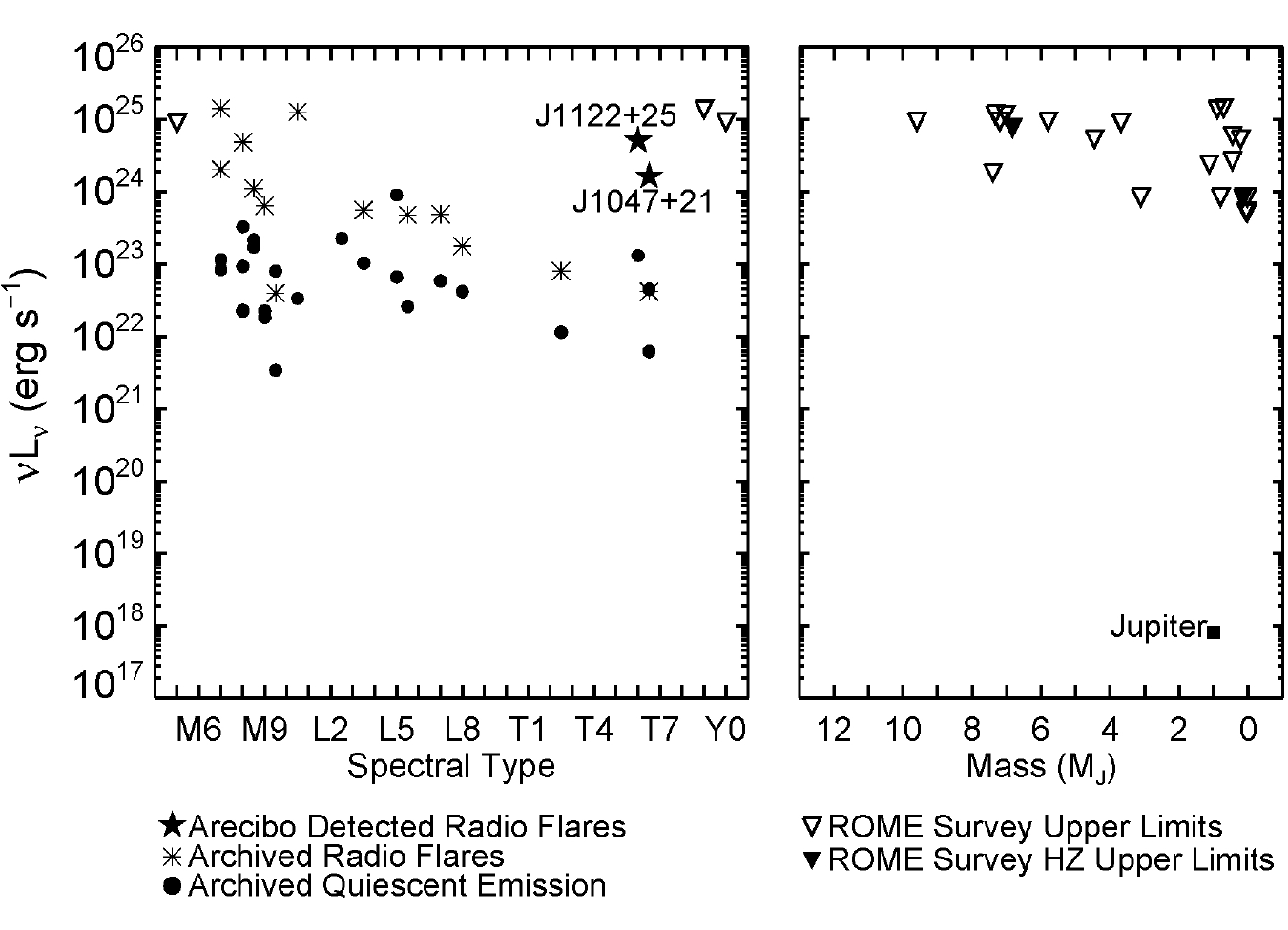}
	\caption{Logarithmic plots of $\nu L_{\nu}$ {\bf} isotropic radio luminosity detection limits for surveyed substellar companions versus spectral type (left panel) or exoplanet mass (right panel), as compared to previously detected radio-emitting UCDs \citep{bur15,rou16b,rou17a,wb15,wil14,wil15,wil17,zic19,ric20}.  The two AO discoveries of the coolest known radio-flaring brown dwarfs near the brown dwarf-exoplanet boundary are marked by filled stars \citep{rou12,rou16a}.  HD 114762 Ab appears at upper left with an estimated spectral type of M5.  HD 106252 b ($\sim$T9) and HD 38529 c ($\sim$Y0) appear in the left panel at upper right.  My earlier radio detection limits for the four young exoplanets of the HR 8799 system are graphed based on their estimated masses (Table 3), as opposed to their spectral types \citep{opp13,rou13}.  Detection limits from exoplanets surveyed in ROME III \citep{rou23} are included in Table 3 and graphed here as well.  Two surveyed exoplanets are within the HZs of their host stars: clockwise from upper left are HD 10697 b and 55 Cnc f \citep{kan12}.  The Jovian auroral decametric radio emission as measured by \emph{Cassini} anchors the right plot in the lower right corner \citep{zar04}.}
\end{figure}

\end{document}